\documentclass[superscriptaddress,amssymb,amsmath,aps,nofootinbib,prd,twocolumn]{revtex4-1}
\usepackage{graphicx}
\usepackage{xcolor}
\usepackage{hyperref}
\hypersetup{colorlinks=true, linkcolor=blue, citecolor=blue}
\usepackage{placeins}
\usepackage{soul}

\begin{document}
\title{Appearance and disappearance of thermal renormalons}
\author{E. Cavalcanti}
\email[]{erich@cbpf.br}
\affiliation{Centro Brasileiro de Pesquisas F\'{\i}sicas/MCTI, Rio de Janeiro, RJ, Brazil}
\author{J. A. Louren\c{c}o}
\email[]{jose.lourenco@ufes.br}
\affiliation{Universidade Federal do Esp\'{i}rito Santo, Campus S\~ao Mateus, ES, Brazil}
\author{C.A. Linhares}
\email[]{linharescesar@gmail.com}
\affiliation{Instituto de F\'{\i}sica, Universidade do Estado do Rio de Janeiro, Rio de Janeiro, RJ, Brazil}
\author{A. P. C. Malbouisson}
\email[]{adolfo@cbpf.br}
\affiliation{Centro Brasileiro de Pesquisas F\'{\i}sicas/MCTI, Rio de Janeiro, RJ, Brazil}

\begin{abstract}
We consider a scalar field model with a $g \phi_4^4$ interaction and compute the mass correction at next-to-leading order in a large-$N$ expansion to study the summability of the perturbative series. It is already known that at zero temperature this model has a singularity in the Borel plane (a ``renormalon"). We find that a small increase in temperature adds two countable sets both with an infinite number of renormalons. For one of the sets the position of the poles is thermal independent and the residue is thermal dependent. In the other one both the position of poles and the residues are thermal dependent.
If we consider the model at extremely high temperatures, however, one observes that all the renormalons disappear and the model becomes Borel summable. 
\end{abstract}


\maketitle


\section{Introduction}

The understanding of strongly-coupled systems remains one of the major challenges in particle physics and requires the knowledge of the nonperturbative regime of Quantum Chromodynamics (QCD), the currently accepted theory of strong interactions.

Also, in the realm of condensed matter physics, systems involving strongly coupled particles (fermions, for instance) fall, in principle, outside the scope of perturbation theory. However, apart from some simple models, nonperturbative solutions are very hard to be found, which led along the years to attempts to rely in some way on perturbative methods (valid in general for weak couplings) to get some results in strong-coupling regimes~\cite{LeGuillou1990,Aniceto:2013fka,Grassi:2014cla,Cherman2015,Dorigoni2014,Bender1999}.

It is broadly discussed in the literature whether nonperturbative solutions in field theory can or cannot be recovered from a perturbative expansion. In any case, a procedure is needed to make sense out of the perturbative series. In fact, often the perturbative expansions are asymptotic rather than convergent. Actually, we remember that the perturbative series can be viewed just as a representation of the exact solution and if we want to obtain information about the nonperturbative solution from its perturbative representation some summation technique must be implemented.~\cite{LeGuillou1990,Aniceto:2013fka, Grassi:2014cla, Cherman2015, Dorigoni2014,Bender1999}

One of the most employed of these procedures is to investigate, after perturbative renormalization has been performed, the so-called \textit{Borel summability} of a theory, for a brief introduction see Refs.~\cite{RivasseauBook1,TheBook} and for a complete review on the subject see Ref.~\cite{Beat1981}
. If we start with an asymptotic series, its Borel transform defines a new series that can be convergent. The representation of the nonperturbative result can be obtained by an inverse Borel transform, essentially a Laplace transform, which requires a contour integration in the complex Borel plane in order to be properly defined. This procedure allows one to gain access to the correct nonperturbative solution in many situations~\cite{Grassi:2014cla}. More precisely: if we take a theory characterized by an already renormalized coupling constant $g$ and consider a given quantity $F(g)$ given by a formal series (perhaps asymptotically divergent) in $g$, $F(g) = \sum_n a_n g^n$; define its Borel transform $\mathcal{B}(F;b)$ as $\mathcal{B}(F;b)=\sum_n a_n b^n /n!$ and the inverse Borel transform as $\widetilde{F}(g)=1/g \int_0^\infty db\; e^{-b/g} \mathcal{B}(F;b)$. 
It can be easily verified that $\widetilde{F}(g)$ reproduces formally the original series $F(g)$. The interesting point is that even if $F(g)$ is divergent the series $\mathcal{B}(F;b)$ may converge and in this case the inverse Borel transform $\widetilde{F}(g)$ defines a function of $g$ which can be considered in some sense as the sum of the original divergent series $F(g)$.
This ``mathematical phenomenon'' is named \textit{Borel summability} and is a way of giving a meaning to divergent perturbative series. However, it is implicitly assumed the absence of singularities (\textit{renormalons}) at least on the real axis of the Borel plane $b$.

On the other hand, the existence of poles in the complex Borel plane makes the procedure ambiguous and thus ill defined. These difficulties are essentially of a nonperturbative nature. Recent developments have investigated this issue considering models for which exact results are known, so that we can be sure that the inverse Borel transform or other techniques can give information about the nonperturbative behavior~\cite{Grassi:2014cla,Cherman2015a}. This quest introduces the resurgence technique, developed by \'Ecalle in a different context~\cite{Ecalle1981} which has led to many applications~\cite{Shifman:2013uka,Beneke:1998ui,Cherman2015,Cherman2015a, Misumi2015,Aniceto2015,Dunne:2012ae}; for a review see Refs.~\cite{Dorigoni2014,Aniceto2018,booksauzin}.  

As pointed out in Shifman's review article~\cite{Shifman:2013uka}, the study of renormalons is also important from a phenomenological perspective, as one needs, for instance, to know the solution at the nonperturbative level to obtain an estimate of the heavy-quark mass~\cite{Beneke1994}.

Recently, the investigation of renormalons and the subsequent task of the resurgence program to ``cure'' the theories have been done, considering compactified theories such as non-Abelian $SU(N)$ gauge theories on $\mathbb{R}^3 \times \mathbb{S}^1$~\cite{Anber:2014sda} and  the $\mathbb{CP}^{N-1}$ non-linear sigma model on $\mathbb{R}^1 \times \mathbb{S}^1$~\cite{Cherman2015,Shifman:2008ja}. It seems that the renormalon ambiguities are canceled by appropriate contributions from instanton-anti-instanton pairs in these theories~\cite{Cherman2015}. The interest of considering a finite spatial extent $L$ or thermal dependence $\beta=1/T$ arises from the fact that for small $L$ or large $T$ a weakly coupled regime is observed due to asymptotic freedom~\cite{Malbouisson2004,Khanna:2012js}. 

A careful study of renormalons for an $SU(N)$ gauge theory has been made in Ref.~\cite{Anber:2014sda}. In it the absence of renormalons is discussed when one introduces a finite small length $L$. Another proposal~\cite{Shifman:2013uka} is to study 2D models with symmetries such as $O(N)$, $CP^{N-1}$, where the small-length regime becomes a quantum mechanical problem.

In the present article, we investigate the behavior of a scalar field model with $O(N)$ symmetry in four dimensions. Our main concern is the careful investigation of the renormalon poles and residues at next-to-leading order in the large-$N$ expansion. This $1/N$-expansion allows to resum a class of diagrams (usually called \textit{ring} diagrams or \textit{necklaces}) that generates the renormalon contribution. Following recent literature, we investigate the role of a compactification parameter, here taken as introducing a temperature dependence.
First, we review the behavior at zero temperature and find the existence of two renormalons. At small temperatures we observe that the system develops a countable set with an infinite number of renormalons that can be separated into two classes: renormalons without thermal poles but that can have thermal residues and renormalons with thermal poles. A further increase in temperature is found to imply the disappearance of renormalons.


\section{Scalar field model and resummation}

\begin{figure}[h]
	\centering
	\includegraphics[width=0.9\linewidth]{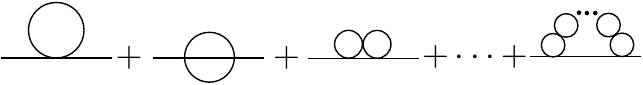}
	\caption{Sum over the class of necklace diagrams. The case without any pearl is the usual `tadpole' (first diagram), the special case with just one pearl is the usual `sunset' diagram (second diagram).}
	\label{fig:Colares}
\end{figure}

We are mainly interested in computing corrections to the field mass in a scalar theory with coupling $g \phi^4$. The full propagator $G$ is given by
\begin{equation}
G = G_0 \sum_{k=0}^\infty (\Sigma G_0)^k \equiv \frac{G_0}{1-\Sigma G_0},
\end{equation}
\noindent where $\Sigma$ is the sum of all 1PI (one-particle irreducible) diagrams built with the free propagator $G_0$. Or, if we establish $\Sigma$ using the full propagator $G$ (a recurrence relation) then, to avoid double counting, it is necessary to consider just the 2PI diagrams. We use a set of 2PI diagrams known as \textit{necklace} or \textit{ring} diagrams as ilustrated in figure~\ref{fig:Colares}.

Therefore, at an unspecified spacetime dimension $D$ a necklace with $(k-1)$-\textit{pearls} is given by $R_k(p)$, 
\begin{equation}
R_k(p) = -g \int \frac{d^D \ell}{(2\pi)^D} \frac{1}{(p-l)^2+M^2} \left[-\frac{g}{2} B(\ell)\right]^{k-1},
\end{equation}
\noindent where
\begin{equation}
B(\ell) = \int \frac{d^D q}{(2\pi)^D} \frac{1}{q^2 + M^2} \frac{1}{(q+l)^2 + M^2}
\end{equation}
\noindent stands for each \textit{pearl}~\cite{LeGuillou1990,zinnjustin2002}.

Thus, taking into account necklaces with all numbers of pearls we obtain the full correction
\begin{equation}
\Sigma = \sum_{k=1}^{\infty} R_k(p). \label{Eq:OverallSum}
\end{equation}
The subsequent analysis of this expression intends to verify whether the series representation is or is not Borel summable. This is entirely dependent on the behavior of $R_k(p)$ with respect to the summation index $k$.

Now that we have introduced the general idea, let us investigate the thermal dependence in detail. By making a compactification in imaginary time we introduce the inverse temperature $\beta=1/T$. With this, the expression for each pearl is modified to
\begin{multline}
B(\ell,\omega_m) = \frac{1}{\beta} \sum_{n\in \mathbb{Z}} \int \frac{d^{D-1}q}{(2\pi)^{D-1}} \frac{1}{q^2 + \omega_n^2 +  M^2}\\ \times \frac{1}{(q+l)^2 + (\omega_n-\omega_m)^2 + M^2},
\label{Eq:Bubble}
\end{multline}
\noindent where $\omega_n = 2\pi n T$ is the frequency related to the $(D-1)$-dimensional momentum $q$, while the necklaces become
\begin{multline}
R_k(p,\omega_o) = \frac{(-g)^k}{2^{k-1} \beta}\\ \times \sum_{m\in \mathbb{Z}} \int \frac{d^{D-1}\ell}{(2\pi)^{D-1}} \frac{B^{k-1}(\ell,\omega_m)}{(p-l)^2 + (\omega_o-\omega_m)^2 + M^2},\label{Eq:Necklace}
\end{multline}
\noindent where $\omega_m$ and $\omega_o$ are the frequencies related, respectively, to the loop momentum $\ell$ and the external momentum $p$.

Then, Eq.~\eqref{Eq:Bubble} can be treated by using the Feynman parametrization, integrating over the momenta $q$ and identifying the infinite sum as an Epstein-Hurwitz zeta function $Z^{X^2}(\beta;\nu)$ defined by
\begin{equation}
Z^{X^2}(\beta;\nu) =  \sum_{m\in \mathbb{Z}} \frac{1}{\left(\omega_m^2+X^2\right)^{\nu}}.
\label{Eq:Zeta}
\end{equation}
We now perform the analytic expansion of the Epstein-Hurwitz zeta function to whole complex $\nu$ plane~\cite{Elizalde:1996zk}, which allows to rewrite Eq.~\eqref{Eq:Bubble} as 
\begin{multline}
\hspace{-5mm}	B(\ell,\omega_m) =\hspace{-1mm} \frac{\Gamma\left(2-\frac{D}{2}\right)}{(4\pi)^{\frac{D}{2}}} \hspace{-2mm}\int_0^1\hspace{-2mm} dz \left[M^2 \hspace{-1mm}+ (\ell^2 + \omega_m^2)z(1-z)\right]^{-2+\frac{D}{2}}\\
	+ \frac{1}{(2\pi)^{\frac{D}{2}}} \sum_{n\in\mathbb{N}^\star} \int_0^1 dz \frac{(n \beta)^{2-\frac{D}{2}}\cos\left[2 \pi n m (1-z)\right]}{\left[M^2 + (\ell^2 + \omega_m^2)z(1-z)\right]^{\frac{2-\frac{D}{2}}{2}}} \\\times K_{2-\frac{D}{2}} \left[n \beta \sqrt{[M^2 + (\ell^2 + \omega_m^2)z(1-z)}\right],
\end{multline}
\noindent where $K_\nu(x)$ is the modified Bessel function of the second kind.

Considering the case where $D=4-2\varepsilon$, we get
\begin{multline}
	B(\ell,\omega_m) = B_0(\ell,\omega_m) + B_\beta(\ell,\omega_m) \\= \frac{\Gamma\left(\varepsilon\right)}{(4\pi)^{2}} \int_0^1 dz \left\{1- \varepsilon \ln \left[M^2 + (\ell^2 + \omega_m^2)z(1-z)\right]\right\}\\
	+ \frac{1}{(2\pi)^2} \sum_{n\in\mathbb{N}^\star} \int_0^1 dz \cos\left[2 \pi n m (1-z)\right] \\ \times K_{0} \left[n \beta \sqrt{[M^2 + (\ell^2 + \omega_m^2)z(1-z)}\right].
    \label{Eq:FullBubble}
\end{multline}

The temperature-independent component $B_0$ is standard and well known~\cite{zinnjustin2002},
\begin{multline}
B_0(\ell,\omega_m) = - \frac{1}{(4\pi^2)} \Bigg\{ \ln\frac{M^2}{\Lambda^2}-2 
+\sqrt{1+\frac{4M^2}{\ell^2+\omega_m^2}} \\ \times \ln\left[ \frac{1+\sqrt{1+4\frac{M^2}{\ell^2+\omega_m^2}}}{1-\sqrt{1+4\frac{M^2}{\ell^2+\omega_m^2}}}\right]
\Bigg\}.\label{Eq:BubbleStandard}
\end{multline} 
\noindent For high values of the momentum $\ell$ we have the asymptotic expression 
\begin{equation}
B_0(\ell,\omega_m) \sim - \frac{1}{(4\pi^2)} \ln \frac{\ell^2+\omega_m^2+M^2}{M^2}. \label{Eq:BubbleT0Asymp}
\end{equation}
\noindent However, we do not have a solution for the term $B_\beta$ for all temperatures. In sections \ref{Sec:LowT} and \ref{Sec:HighT} we respectively investigate the regimes of low and high temperatures.


\section{A first glance : renormalon at $T=0$ \label{Sec:ZeroT}}
In this section we consider the special case of zero temperature. In this situation the only contribution to the pearl diagram comes from the $B_0$ component. To obtain a treatable expression to the necklace diagrams, we consider the expansion for high values of the momentum $\ell$, Eq.~\eqref{Eq:BubbleT0Asymp}, at zero temperature 
\begin{equation}
B(\ell,\omega_m) \overset{T = 0}{\sim} -\frac{1}{(4\pi)^2} \ln \frac{\ell^2+M^2}{M^2}.
\label{Eq:Bubble_NullT}
\end{equation}

At this point we recall that the standard approximation is to consider the leading behavior in the momentum $\ell$, that is, $\ln (\ell^2+M^2) \approx \ln \ell^2$. Here, we avoid this particular approximation and explore the consequences of keeping the exact term $\ln (\ell^2+M^2)$. Let us return to the necklace diagrams. We employ a BPHZ procedure defining a renormalized necklace $\hat R_k(p)$
\begin{equation}
\hat R_k (p) = R_k (p) - R_k (0) - p^2 \frac{\partial}{\partial p^2} R_k (p) \Bigg|_{p=0}.
\end{equation}
\noindent We shall drop the hat unless it becomes important to distinguish between the renormalized $\hat R_k (p)$ and non renormalized $R_k (p)$ necklaces.

As can be noted, this affects only the $p$-dependent propagator in the zero-temperature version of Eq.~\eqref{Eq:Necklace}. The procedure is equivalent to perform the substitution
\begin{multline}
\frac{1}{p^2+\ell^2+M^2} \overset{BPHZ}{\longrightarrow} \frac{p^4}{(\ell^2+M^2)^2(p^2+\ell^2+M^2)} \\ \approx
\begin{cases}
\frac{p^4}{(\ell^2+M^2)^3}, \quad\text{low}\; p;\\
\frac{p^2}{(\ell^2+M^2)^2}, \quad\text{high}\; p,
\end{cases}
\label{Eq:BHPZp}
\end{multline}
\noindent where the standard \textit{naive} expansion $(p-l)^2\approx p^2+l^2$ is assumed.

In a low-$p$ expansion, then, for $T=0$ and small values of $p$, the integral to be solved to obtain the necklace expression is
\begin{equation}
R_k(p) \sim -g p^4 \left( \frac{g}{2(4\pi)^2} \right)^{k-1} \int \frac{d^{4}\ell}{(2\pi)^{4}} \frac{\left( \ln \frac{\ell^2+M^2}{M^2}\right)^{k-1}}{(\ell^2+M^2)^3}. \label{Eq:RkT0}
\end{equation}
\noindent To solve it we first perform the integral over the solid angle ($\Omega_4 = 2 \pi^2$) and then reorganize the result by making the change of variables $\ell^2+M^2 = M^2 e^t$, that is,
\begin{equation}
R_k(p) \sim - \frac{g p^4}{16\pi^2 M^2} \left( \frac{g}{2(4\pi)^2} \right)^{k-1} \int_0^\infty dt (e^{-t}-e^{-2t}) t^{k-1}.
\end{equation}
\noindent At this point we can clearly identify the presence of two gamma functions, so that
\begin{multline}
R_k(p) \sim - \frac{g p^4}{16\pi^2 M^2} 
\Bigg\{ (k-1)! \left( \frac{g}{2(4\pi)^2} \right)^{k-1}
\\- (k-1)! \left( \frac{g}{4(4\pi)^2} \right)^{k-1}
\Bigg\}.
\end{multline}

We can finally return to the sum over all contributions Eq.~\eqref{Eq:OverallSum},
\begin{equation}
\Sigma \sim -\frac{2 \widetilde{g} p^2}{M^2} \left\{
\sum_{k=1}^{\infty} (k-1)!\widetilde{g}^{k-1}
-
\sum_{k=1}^{\infty} (k-1)! \left(\frac{\widetilde{g}}{2}\right)^{k-1}
\right\},
\end{equation}
\noindent where we have defined $\widetilde{g} = g/(2 (4\pi)^2)$. Both sums are divergent, but we can try to make sense of them by defining a Borel transform,
\begin{align}
\mathcal{B}(\Sigma;y) &\sim -\frac{2 \widetilde{g} p^2}{M^2} \left\{
\sum_{k=1}^{\infty} (\widetilde{g} y)^{k-1}
- \sum_{k=1}^{\infty} \left(\frac{\widetilde{g}y}{2}\right)^{k-1}
\right\} \nonumber\\
&= -\frac{2 \widetilde{g} p^2}{M^2} \left\{
\frac{1}{1-\widetilde{g} y}
- \frac{1}{1-\widetilde{g}y/2}
\right\}
\end{align}

We then obtain two poles on the real positive axis of the Borel plane at $y = 1/\widetilde{g}, 2/\widetilde{g}$. These poles (renormalons) introduce problems to compute the inverse Borel transform.

In the standard procedure, see Ref.~\cite{zinnjustin2002}, Eq.~\eqref{Eq:RkT0} is solved for very large $\ell$, which is justified as this is the relevant region to get the asymptotic behavior for the $k$ index. This means that the approximation $\left[\ln (\ell^2+M^2)/M^2\right]^{k-1} / (\ell^2+M^2)^3 \approx \left( \ln \ell^2\right)^{k-1}/\ell^6$ is employed. Therefore, only the first pole is found (at $y=1/\widetilde{g}$) while the second pole is hidden. When $\widetilde{g}$ is very small this could be justified as $2/\widetilde{g}$ being very far from the origin.

\section{Appearance of thermal renormalons (low temperatures) \label{Sec:LowT}}
 
For low but finite temperatures, we can use the asymptotic representation of the modified Bessel function of the second kind  $K_0(z) \sim e^{-z} f(z)$, so that the thermal component of the pearl \eqref{Eq:FullBubble} becomes
\begin{equation}
B_\beta(\ell,\omega_m) \sim \frac{1}{(4\pi)^2} \sum_{n\in \mathbb{N}^\star}\frac{8K_0(n\beta M)}{n\beta}\frac{1}{\ell^2+\omega_m^2}.
\end{equation}

Using the above equation for $B_\beta$ and the expression for the $T=0$ component, $B_0$ (see Eq.~\eqref{Eq:BubbleT0Asymp}), the quantity $B = B_0 + B_\beta$ can be written in the low-temperature regime as
\begin{equation}
B(\ell,\omega_m) \sim -\frac{1}{(4\pi)^2} \left[ \ln \frac{\ell^2+\omega_m^2+M^2}{M^2} -\frac{A(\beta)}{\ell^2+\omega_m^2} \right], \label{Eq:Bubble_LowT}
\end{equation}
\noindent where
\begin{equation}
A(\beta) = \frac{1}{(4\pi)^2} \sum_{n\in \mathbb{N}^\star}\frac{8K_0(n\beta M)}{n\beta}
\end{equation}
\noindent stores information about the dependence on the temperature.

We then replace the expression in Eq.\eqref{Eq:Bubble_LowT} into Eq.\eqref{Eq:Necklace}, employ the BHPZ procedure and use a low-$p$ expansion as in Eq.\eqref{Eq:BHPZp},
\begin{multline}
R_k(p,\omega_o) = -g (p^2+\omega_o^2)^2 \left( \frac{g}{2(4\pi)^2} \right)^{k-1} \frac{1}{\beta} \\ \times\sum_{m\in \mathbb{Z}} \int \frac{d^{3}\ell}{(2\pi)^{3}} \frac{\left( \ln\frac{\ell^2+\omega_m^2+M^2}{M^2} -\frac{A(\beta)}{\ell^2+\omega_m^2} \right) ^{k-1}
	}{(\ell^2+\omega_m^2 +M^2)^3}. \label{Eq:RkThermal}
\end{multline}
\noindent So, integrating over the solid angle and expanding the binomial, we get
\begin{multline}
R_k(p,\omega_o) = -\frac{g (p^2+\omega_o^2)^2}{2\pi^2} \left( \frac{g}{2(4\pi)^2} \right)^{k-1} \\ \times \frac{1}{\beta} \sum_{m\in \mathbb{Z}} \sum_{i=0}^{k-1} {k-1\choose i}  (-A(\beta))^i\\ \times\int_0^\infty d\ell\; \ell^2 \frac{\ln^{k-1-i} [(\ell^2+\omega_m^2+M^2)/M^2] 
}{(\ell^2+\omega_m^2 +M^2)^3} \left(\frac{1}{\ell^2+\omega_m^2}\right)^i.
\end{multline}
\noindent We reorganize the above expression in a more convenient way to compute the sum over the Matsubara frequencies. The denominator is treated by employing a Feynman parametrization and the logarithm in the numerator is expanded in powers of $\omega_m^2/(\ell^2+M^2)$, which is justified by an asymptotic behavior in $\ell$ assuring that $m/\ell < 1$. This allows to rewrite the above equation in the form
\begin{widetext}
\begin{multline}
R_k(p,\omega_o) = -\frac{g (p^2+\omega_o^2)^2}{2\pi^2} \left( \frac{g}{2(4\pi)^2} \right)^{k-1} \sum_{i=0}^{k-1} {k-1\choose i}  (-A(\beta))^i \int_0^1 dz \frac{\Gamma(3+i)z^2 z^{i-1}}{\Gamma(3)\Gamma(i)}\\
\times\Bigg\{\int_0^\infty d\ell\; \ell^2 \ln^{k-i-1} \frac{\ell^2+M^2}{M^2}
\frac{1}{\beta} \sum_{m\in \mathbb{Z}} \frac{1}{\left(\ell^2+\omega_m^2+M^2z\right)^{3+i}}
+
\int_0^\infty d\ell\; \ell^2 \frac{\ln^{k-i-2} \frac{\ell^2+M^2}{M^2}}{\ell^2+M^2}
\frac{1}{\beta} \sum_{m\in \mathbb{Z}} \frac{(k-i-1)\omega_m^2}{\left(\ell^2+\omega_m^2+M^2z\right)^{3+i}}
+
\mathcal{O}(\omega_m^4)
\Bigg\}. \label{Eq:Rkfull}
\end{multline}
\end{widetext}

Although we could use this complete expression, this is unnecessary. It can be shown, after a lengthy computation, that the relevant information (poles in the Borel plane) can already be obtained by using the following approximation,
\begin{multline}
R_k(p,\omega_o) \approx -\frac{g (p^2+\omega_o^2)^2}{2\pi^2} \left( \frac{g}{2(4\pi)^2} \right)^{k-1} \\ \times\sum_{i=0}^{k-1} {k-1\choose i}  (-A(\beta))^i
 \Bigg\{\int_0^\infty d\ell\; \ell^2 \ln^{k-1-i} \frac{\ell^2+M^2}{M^2} \\
\times\frac{1}{\beta}Z^{\ell^2+M^2}(\beta;3+i)
\Bigg\}, \label{Eq:Rkapprox}
\end{multline}
\noindent where $Z^{X^2}(\beta;\nu)$ is the Epstein-Hurwitz zeta function defined in Eq.~\eqref{Eq:Zeta}. The contributions of order $\mathcal{O}(\omega_m^2)$ do not modify the position of the poles and only change their residues. Moreover, for large values of $k$ the integration of the expression over the Feynman parameter $z$ is asymptotically equal to the expression without the Feynman parameters. To avoid a tedious calculation we do not exhibit in this article the step-by-step of this process.

Taking the approximation in Eq.~\eqref{Eq:Rkapprox} and considering again the analytic expansion of the Epstein-Hurwitz zeta function to the whole complex $\nu$ plane, we get
\begin{widetext}
\begin{multline}
R_k(p,\omega_o) \approx 
-\frac{g (p^2+\omega_o^2)^2}{2\pi^2} \left( \frac{g}{2(4\pi)^2} \right)^{k-1} \sum_{i=0}^{k-1} {k-1\choose i}  (-A(\beta))^i \int_0^\infty d\ell \; \ell^2\ln^{k-1-i} \frac{\ell^2+M^2}{M^2}\\ \times
\frac{1}{\sqrt{4\pi} \Gamma\left(3+i\right)} \left\{
\frac{\Gamma\left(\frac{5}{2} +i\right)}{(\ell^2+M^2)^{\frac{5}{2}+i}}
+\frac{4}{2^{\frac{5}{2}+i}} \sum_{n\in \mathbb{N}^\star} \left(\frac{n \beta}{\sqrt{\ell^2+M^2}}\right)^{\frac{5}{2}+i} K_{\frac{5}{2}+i} (n \beta \sqrt{\ell^2+M^2})
\right\}.
\end{multline}
\end{widetext}
\noindent Since $i$ is an integer, the modified Bessel function of the second kind has a half-integer order, which has the series representation~\cite{Gradshteyn}
\begin{multline}
K_{\frac{5}{2}+i} (n \beta \sqrt{\ell^2+M^2}) = \sqrt{\frac{\pi}{2}} \sum_{j=0}^{i+2} \frac{(j+i+2)!}{j!(i+2-j)!}\\ \times \frac{1}{2^j(n \beta)^{j+\frac{1}{2}}} \frac{e^{-n \beta \sqrt{\ell^2+M^2}}}{(\ell^2+M^2)^{j+\frac{1}{2}}}.
\end{multline}
So, the remaining integrals are given by
\begin{multline}
R_k(p,\omega_o) = 
-\frac{g (p^2+\omega_o^2)^2}{4\pi^{5/2}} \left( \frac{g}{2(4\pi)^2} \right)^{k-1} \sum_{i=0}^{k-1} {k-1\choose i} \\  \times \frac{(-A(\beta))^i}{(2+i)!}
\Bigg\{
\Gamma\left(\frac{5}{2} +i\right) \int_0^\infty d\ell\; \ell^2 \frac{\ln^{k-1-i} \frac{\ell^2+M^2}{M^2}}{(\ell^2+M^2)^{\frac{5}{2}+i}} \\
+\frac{\sqrt{\pi}}{2^{1+i}} \sum_{n\in \mathbb{N}^\star} \sum_{j=0}^{i+2} \frac{(j+i+2)!}{j!(i+2-j)!} \frac{1}{2^j(n \beta)^{j-i-2}}\\
\times \int_0^\infty d\ell\; \ell^2 \frac{e^{-n \beta \sqrt{\ell^2+M^2}}}{(\sqrt{\ell^2+M^2})^{2j+i+\frac{7}{2}}} \ln^{k-1-i} \frac{\ell^2+M^2}{M^2} \Bigg\}.\label{Eq:Rktointegrate}
\end{multline}

The first integral in the preceding equation can be solved as in the zero-temperature case (see Sec.~\ref{Sec:ZeroT}) by the change of variables $\ell^2+M^2 = M^2 e^t$. One must note that $\sqrt{e^t-1}$ has an upper bound $\sqrt{e^t}$ that is also its asymptotic value for large values of the momentum $t$ (which means also large values of the index $k$). Then, we can use that $\sqrt{e^t-1} \lesssim \sqrt{e^t}$ to simplify the integral
\begin{multline}
\mathcal{I}_1=\int_0^\infty d\ell\; \ell^2 \frac{\ln^{k-1-i} \frac{\ell^2+M^2}{M^2}}{(\ell^2+M^2)^{\frac{5}{2}+i}}
\\= \frac{1}{2 M^{2+2i}}
 \int_0^\infty dt\; t^{k-i-1} \sqrt{e^t-1}
 e^{-t\left(\frac{3}{2}+i\right)}
\lesssim \frac{1}{2 M^{2+2i}} \\
\times \int_0^\infty dt\; t^{k-i-1}
e^{-t\left(1+i\right)}
=
\frac{(k-i-1)!}{2 M^{2+2i}} 
\frac{1}{\left(1+i\right)^{k-i}}.
\label{Eq:Rk_firstint}
\end{multline}

For the second integral in Eq.~\eqref{Eq:Rktointegrate} we make the change of variables $\ell^2+M^2 = M^2 r^2$ so that we obtain
\begin{multline}
\mathcal{I}_2=\int_0^\infty d\ell\; \ell^2 \frac{e^{-n \beta \sqrt{\ell^2+M^2}}}{(\sqrt{\ell^2+M^2})^{2j+i+\frac{7}{2}}} \ln^{k-1-i} \frac{\ell^2+M^2}{M^2}
\\=
2^{k-i-1} M^{3-2(j+i/2+7/4)}\int_1^\infty dr\; \sqrt{r^2-1} \frac{e^{-n \beta r} \ln^{k-i-1} r}{r^{2(j+i/2+7/4)-1}}\\
\lesssim 2^{k-i-1} M^{3-2(j+i/2+7/4)}\int_1^\infty dr\; \frac{e^{-n \beta r} \ln^{k-i-1} r}{r^{2(j+i/2+7/4)-2}}. \label{Eq:Rk_secondint}
\end{multline}
\noindent Once  more, we used that $\sqrt{r^2-1} \lesssim r$, which considerably simplify the integral and allows it to be identified as the Milgram generalization of the integro-exponential function whose asymptotic behavior is known~\cite{Milgram1985}
\begin{multline}
E_s^\alpha (z) = \frac{1}{\Gamma(\alpha+1)} \int_1^\infty dt \frac{(\ln t)^\alpha e^{-zt}}{t^s} \\\overset{\text{Re} z \rightarrow \infty}{\sim} \frac{e^{-z}}{z^{\alpha+1}} \left[ 1 - \frac{(\alpha+1)(\alpha+2s)}{2z} + \ldots \right]. \label{Eq:IntegroExponential}
\end{multline}

Hence, substituting Eqs.~\eqref{Eq:Rk_firstint} and \eqref{Eq:Rk_secondint} into Eq.~\eqref{Eq:Rktointegrate}, and using the asymptotic behavior of the generalized integro-exponential, Eq.~\eqref{Eq:IntegroExponential}, we have
\begin{multline}
	R_k(p,\omega_o) \lesssim
	-\frac{g (p^2+\omega_o^2)^2}{4\pi^{5/2}} \left( \frac{g}{2(4\pi)^2} \right)^{k-1} \sum_{i=0}^{k-1} {k-1\choose i} \\ \times \frac{(-A(\beta))^i(k-i-1)!}{(2+i)!} 
	\Bigg\{
	\Gamma\left(\frac{5}{2} +i\right) 
	\frac{1}{2 M^{2+2i}} 
	\frac{1}{\left(1+i\right)^{k-i}}
	 \\+
	\sqrt{\pi} \sum_{n\in \mathbb{N}^\star} \sum_{j=0}^{i+2} \frac{(j+i+2)!}{j!(i+2-j)!} \frac{2^{k-2i-j-2}}{(n \beta)^{k+j-2i-2}}
	\frac{e^{-n \beta}}{M^{2j+i+1/2}}	
	\Bigg\}.
\end{multline}

Now, let us focus attention on the $k$-dependence. The previous equation can then be rewritten as
\begin{multline}
	R_k \lesssim
	\gamma_1 \widetilde{g}^{k-1} (k-1)! \sum_{i=0}^{k-1}
	\Bigg[
	\frac{\gamma_{2,i}(\beta,M)}{\left(1+i\right)^{k-1}}	
	\\+
	\sum_{n\in \mathbb{N}^\star} \frac{2^{k-1}\gamma_{3,i,n}(\beta,M)}{(n \beta)^{k-1}}
	\Bigg],
\end{multline}
\noindent where we have defined,
\begin{subequations}
\begin{align}
\gamma_1 &= -\frac{g (p^2+\omega_o^2)^2}{4\pi^{5/2}},\\
\widetilde{g} &= \frac{g}{2(4\pi)^2},\\
\gamma_{2,i}(\beta,M) &= \frac{(-A(\beta))^i}{(2+i)!i!} \Gamma\left(\frac{5}{2} +i\right) 
\frac{\left(1+i\right)^{i-1}}{2M^{2+2i}}, \\
\gamma_{3,i,n}(\beta,M) &= \sum_{j=0}^{i+2} \frac{(-A(\beta))^i}{(2+i)!i!} \sqrt{\pi}  \frac{(j+i+2)!}{j!(i+2-j)!} \nonumber \\& \times \frac{e^{-n \beta}}{M^{2j+i+1/2}} \frac{2^{-2i-j-1}}{(n \beta)^{j-2i-1}}.
\end{align}
\end{subequations}

The sum over all necklaces is then
\begin{multline*}
\mathcal{R} = \sum_{k\in \mathbb{N}^\star} R_k \lesssim
\sum_{k\in \mathbb{N}^\star}\gamma_1 \widetilde{g}^{k-1} (k-1)! \sum_{i=0}^{k-1}
\Bigg[
\frac{\gamma_{2,i}(\beta,M)}{\left(1+i\right)^{k-1}}\\+
\sum_{n\in \mathbb{N}^\star} \frac{2^{k-1}}{(n \beta)^{k-1}} \gamma_{3,i,n}(\beta,M)
\Bigg].
\end{multline*}

The range of summation for the double sum is $0\le i < k < \infty$; we can change the sum ordering and then split the sum over the index $k$ in the form
\begin{equation*}
\sum_{k\in \mathbb{N}^+} \sum_{i=0}^{k-1} f_{i,k}
=
\sum_{i=0}^{\infty} \sum_{k=i+1}^\infty f_{i,k}
=
\sum_{i=0}^{\infty} \left(\sum_{k=1}^\infty f_{i,k} - \sum_{k=1}^{i+1} f_{i,k}\right).
\end{equation*}

The first double sum has the dominant contribution; this can be seen by checking for each value of $i$. Therefore, the relevant contribution is
\begin{multline}
	\mathcal{R} \sim
	\gamma_1 \sum_{i=0}^{\infty}
	\Bigg[
	\gamma_{2,i} \sum_{k=1}^\infty 
	\frac{\widetilde{g}^{k-1} (k-1)! }{\left(1+i\right)^{k-1}}
	\\+
	\sum_{n\in \mathbb{N}^\star} \gamma_{3,i,n} \sum_{k=1}^\infty
	\frac{\widetilde{g}^{k-1} (k-1)!}{(n \beta/2)^{k-1}} 
	\Bigg].
\end{multline}

However, this is not summable due to the presence of the $(k-1)!$. To overcome this difficulty we can employ a Borel transform,
\begin{align}
	\mathcal{B}(\mathcal{R};y)
	&\sim
	\gamma_1 \sum_{i=0}^{\infty}
	\Bigg[
	\gamma_{2,i} \sum_{k=1}^\infty 
	\frac{\widetilde{g}^{k-1} y^{k-1} }{\left(1+i\right)^{k-1}}
	\nonumber \\ &+
	\sum_{n\in \mathbb{N}^\star} \gamma_{3,i,n} \sum_{k=1}^\infty
	\frac{\widetilde{g}^{k-1} y^{k-1}}{(n \beta/2)^{k-1}} 
	\Bigg] \nonumber\\
	&=
	\gamma_1 \sum_{i=0}^{\infty} \Bigg[
	\frac{\gamma_{2,i}(\beta,M)}{1 - \frac{\widetilde{g} y}{1+i}}
	+
	\sum_{n\in \mathbb{N}^\star} \frac{\gamma_{3,i,n}(\beta,M)}{1 - \frac{2\widetilde{g} y}{n\beta}} 
	\Bigg]. \label{Eq:Phi_LowT}
\end{align}

Finally, we see in Eq.~\eqref{Eq:Phi_LowT} the renormalons that appear for low temperatures. There are two different sets of renormalons both with residues that are thermal dependent, respectively $\gamma_{2,i}(\beta,M)$ and $\gamma_{3,i,n}(\beta,M)$. The first set of renormalons was already found in previous works~\cite{Loewe:1999kw}; it is characterized by poles whose position are thermal-independent and they are located along the real axis at positions $(1+i)/\widetilde{g}$ for $i \in \mathbb{N}$. However, the second set of poles, as far as we know, has not yet been reported. These poles are also in the real axis but they are thermal-dependent as they are located at $n \beta / 2 \widetilde{g}$ for $n \in \mathbb{N}^\star$. The existence of this new set seems to be a remarkable enrichment for the model.

We remark that in the limit of extremely small temperatures this new set of renormalons are all very far from the origin and this may justify why they are usually hidden. Therefore, our result can be viewed as a first correction to the standard approach. Furthermore, as we pointed out before, we claim that our approximation in Eq.~\eqref{Eq:Rkapprox} is the sufficient one (at least to describe the poles) and any further corrections shall only change the residues. This means that we have mapped \textit{all} the renormalons that appear at low temperatures.

As a further comment, we remember that in Sec.~\ref{Sec:ZeroT} we show that at zero temperature there is a hidden second pole located at $2/\widetilde{g}$. This does not add any new poles at low temperatures because, as can be easily noted, we already have an infinite set of poles located at $i/\widetilde{g}$ for $i \in \mathbb{N}^\star$.

\section{Disappearance of thermal renormalons (high temperatures) \label{Sec:HighT}}

In this section we explore the regime of very high temperatures. In this situation, to treat Eq.~\eqref{Eq:FullBubble} we can use the following series expansion of the modified Bessel function of the second kind~\cite{Gradshteyn},
\begin{equation}
K_0 (z) = - \ln \frac{z e^\gamma}{2} - \frac{z^2}{4} \ln \frac{z e^{\gamma-1}}{2} + \mathcal{O}(z^4).
\end{equation}

The result is easier to get by assuming from the beginning that $m=0$ (which means that this is the only relevant mode) and recalling the following properties of the Riemann zeta function, $\zeta(s) = \sum_{n\in\mathbb{N}^\star} n^{-s}$,
\begin{subequations}
\begin{align}
\zeta'(s) &= -\sum_{n\in\mathbb{N}^+} \frac{\ln n}{n^s},\\
\zeta(0)&=-1/2,\\
\zeta'(0)&=\ln \sqrt{2\pi}, \\
\zeta(-2k)&=0,\quad \forall k \in \mathbb{N}^\star ,\\
\zeta'(-2k)&=\frac{(-1)^k\zeta(2k+1)(2k)!}{2^{2k+1}\pi^{2k}},\quad \forall k \in \mathbb{N}^\star.
\end{align}
\end{subequations}
\noindent Remembering Eq.~\eqref{Eq:FullBubble}, we then obtain the result
\begin{multline}
B_\beta(\ell,\omega_m) \sim -B_0(\ell,\omega_m) - \frac{1}{8\pi^2} \ln \frac{4\pi T}{\Lambda e^\gamma} \\- \frac{\zeta(3)}{2^7 \pi^4} \left(M^2+\frac{\ell^2}{6}\right)\frac{1}{T^2},
\end{multline}
\noindent revealing that at high temperatures the original contribution from zero temperature is not present anymore. This has a major impact and is responsible for the disappearance of the renormalons. Therefore, we may write
\begin{equation}
B(\ell,\omega_m) \overset{T\rightarrow\infty}{\sim} - \frac{1}{8\pi^2} \ln \frac{4\pi T}{\Lambda e^\gamma} - \frac{\zeta(3)}{2^7 \pi^4} \left(M^2+\frac{\ell^2}{6}\right)\frac{1}{T^2}. \label{Eq:Bubble_HighT}
\end{equation}

If we replace this back into the necklace expression $R_k(p,o)$, in Eq.~\eqref{Eq:RkThermal}, we get 
\begin{multline}
R_k(p,\omega_o) = -g (p^2+\omega_o)^2 \left( \frac{g}{2(4\pi)^2} \right)^{k-1} \frac{1}{\beta}\\ \times \sum_{n \in \mathbb{Z}}\int \frac{d^{3}\ell}{(2\pi)^{3}} \frac{\ln^{k-1}\left( \frac{4\pi T}{\Lambda e^\gamma} \right)}{(\ell^2+\omega_n^2 +M^2)^3}.
\end{multline}
\noindent Since the integration over the internal loop is independent of $k$ we find that 
\begin{equation}
R_k \propto \left(2\tilde{g}\ln \left( \frac{4\pi T}{\Lambda e^\gamma} \right)\right)^{k-1},
\end{equation}
\noindent and, therefore, there is no renormalon in this case. 

The function $\Sigma(g) = \sum_{k=1}^{\infty} R_k(p)$ is Borel summable,
\begin{equation*}
\Sigma \sim \frac{1}{1-2\tilde{g}\ln \left( \frac{4\pi T}{\Lambda e^\gamma} \right)};
\end{equation*}
\noindent it is a meromorphic function of the coupling constant $\widetilde{g}$ having a simple pole at $\widetilde{g} = \left[2\ln \left( 4\pi T / (\Lambda e^\gamma) \right)\right]^{-1}$.


\vspace*{1cm}
\section{Conclusion}
In this article we study the existence of renormalons in a scalar field theory with a $g \phi^4_4$ coupling at next-to-leading order in a large-$N$ expansion. The results in the literature report that there is one renormalon pole at zero temperature (located at $y=1/\widetilde{g}$) and there is an appearance of a countable infinite set of renormalons at low temperatures with the property that the poles are thermal-independent (located at $y=i/\widetilde{g}$, for $i \in \mathbb{N}^\star$). Although, in this article, the standard behavior is reproduced, we also manage to identify the existence of hidden poles, both at zero temperature and at low temperatures. As far as we know, it seems that this fact has not been noted in the literature. Perhaps, these poles where hidden by the approximations used. The extra pole at zero temperature is slightly shifted on the real axis ($y=2/\widetilde{g}$) and can be ignored, as it is done currently in the literature, if the coupling is small enough. At low temperatures, however, there is an entirely new set of renormalons on the real axis that are located at $y = n \beta / 2 \widetilde{g}$ for $n \in \mathbb{N}^\star$. The appearance of renormalons with a small increase in temperature is a remarkable feature of the theory. In this paper we claim that we have mapped all the poles that occur at low temperatures, therefore identifying completely the thermal renormalons that appear. Any further approximation would only improve the value of the residues, but would not modify the number nor the positions of the poles in the Borel plane.

Furthermore, we obtain that at very high temperatures no renormalon singularities occur and the series becomes Borel summable. This seems to indicate that we could think about a ``critical temperature'' where renormalons appear/disappear. This will be the subject of investigation in future work. 

\acknowledgments{The authors thank the Brazilian agency CNPq for partial financial support.}

\bibliography{cleanlibrary}{}
\bibliographystyle{apsrev4-1}

\end{document}